\documentclass{mn2e}
\usepackage{psfig}

\def\ltsima{$\; \buildrel < \over \sim \;$}
\def\lsim{\lower.5ex\hbox{\ltsima}}
\def\gtsima{$\; \buildrel > \over \sim \;$}
\def\gsim{\lower.5ex\hbox{\gtsima}}

\begin{document}

\title[Soft X-ray absorption in Gamma-Ray Bursts]
{Determining the location of Gamma-Ray Bursts through the evolution
of their soft X-ray absorption}

\author[Lazzati \& Perna]
{Davide Lazzati$^1$ \& Rosalba Perna$^{2,3}$ \\ 
$^1$ Institute of Astronomy, University of Cambridge, Madingley Road,
Cambridge CB3 0HA, England \\
$^2$ Harvard Society of Fellows, 78 Mt. Auburn Street, Cambridge, MA
02138\\ 
$^3$ Harvard-Smithsonian Center for Astrophysics, 60 Garden Street,
Cambridge, MA 02138 \\
{\tt e-mail: lazzati@ast.cam.ac.uk}
}

\maketitle

\begin{abstract}
We investigate through dedicated numerical simulations the evolution
of the soft X-ray absorption properties of a cloud surrounding a
gamma-ray burst source. We show that the absorption properties of the
material are strongly modified by the ionization induced by the intense
burst flux. We derive the temporal evolution of the measured column
density as a function of the density and size of the absorbing
medium.  Even if their statistical significance is not extremely
compelling, we find that the detection in several bursts of variable
absorption during the gamma-ray phase can be accounted for if these
bursts are associated to overdense regions in molecular clouds with
properties similar to those of star formation globules. We fit our
model variable column density to the data of GRB~980329 and
GRB~780506, showing that with this method the size, density and
density distribution of the material surrounding a burst can be
measured.
\end{abstract}


\section{Introduction}

An important way to unveil the nature of the Gamma-Ray Burst (GRB)
progenitors is to investigate the properties of the ambient medium
surrounding them.  The most popular models for GRB formation involve
either the collapse of a single massive star or the coalescence of two
compact objects such as two neutron stars or a neutron star and a
black hole (Eichler et al. 1989; Narayan et al. 1992; Woosley 1993)
These models could be constrained by knowing the GRB environment.
Massive stars have very short lives, thus they will explode in
star-forming regions, which are typically characterized by very dense
environments. On the other hand, most merging neutron stars would be
very old and would have typically traveled far from their
birthplace. The scenario of compact merger progenitors could thus be
suggested by medium to low density environments. Recent observations
(see e.g. Kulkarni et al. 1999) for the subclass of long bursts seem
to suggest their association with star-forming regions.

Within the context of GRB progenitors involving the collapse of a
massive star, several proposals have been made: the so-called
``hypernova'' scenario (Paczy\'nski 1998; MacFadyen \& Woosley 1998)
predicts a simultaneous explosion of the burst with a Type Ic
supernova, while the ``Supranova'' model by Vietri \& Stella (1998)
predicts a two-step explosion, in which the burst onset follows a
supernova explosion by several months.  While the rebrightening of the
optical afterglows detected in several bursts $\sim30$ days after the
burst explosion (Bloom et al. 1999; Reichart 1999; Galama et al. 2000)
is suggestive of the first scenario, the detection of iron emission
and absorption features in the X-ray emission of the bursts and
afterglows (Amati et al. 2000; Piro et al. 2000) can be better
explained with a two-step explosion (Lazzati et al. 2001a).
Independent tests of these different scenarios are therefore needed. A
study of the density distribution in the close environment of the
bursts can be a powerful diagnostic of the type of progenitors. In
fact, if the burst explodes simultaneously with the supernova, the
photons propagate through the pre-explosion stellar wind [$n(r)
\propto r^{-2}$], while in the two-step scenario a high density metal
enriched supernova remnant is expected to surround the burst explosion
site (Lazzati et al. 1999). 

Various methods have been proposed to investigate the properties of
the GRB environment. It has been shown that the effect of the
surrounding material on the propagation of photons can be a probe of
the material itself through the detection of absorption features and
their time evolution (Perna \& Loeb 1998; Meszaros \& Rees 1998;
B\"ottcher et al. 1999; Ghisellini et al. 1999; Lazzati et
al. 2001abc; Ramirez-Ruiz et al. 2001a).  In X-rays, it is
particularly interesting the possibility to detect iron $K_\alpha$
absorption edges (Ghisellini et al. 1999; B\"ottcher et al. 1999, Weth
et al. 2000) and their time evolution (Lazzati et al. 2001c;
B\"ottcher et al. 2001). In the optical, the evolution of the
equivalent width of absorption features has been investigated by Perna
\& Loeb (1998).  On the other hand, modelling the afterglow spectral
and intensity evolution in the framework of the external shock
synchrotron model can also constrain the type of environment (Wijers
\& Galama 1999; Berger et al. 2000; Panaitescu \& Kumar 2001,
Ramirez-Ruiz et al. 2001b). In the first case, however, the search
for the signatures of the material is made difficult by the low
signal-to-noise ratio of the lines, while the second procedure heavily
relies on a particular model, and hence depends on its assumptions.

In this paper, rather than considering the feature produced by a
particular transition, we consider the continuum soft X-ray
absorption caused by the photoionization of low-intermediate atomic
number ($Z$) atoms along the line of sight to the burst (Morrison \&
McCammon 1983). This absorption is usually quantified by means of the
quantity $N_H$, the column density of neutral material with solar
metallicity that would cause the observed absorption (in the
simulations we use abundances from Anders \& Ebihara 1982). The
presence of $N_H$ is generally much easier to detect with respect to a
Fe $K_\alpha$ absorption edge. In fact, the opacity of iron at
$7.1$~keV is less than one hundredth of the opacity at 1~keV of a
material with solar iron abundance (Morrison \& McCammon 1983). On the
other hand, the precise measurement of the absolute $N_H$ value can be
model dependent.  In fact, the $N_H$ measurement is performed by
computing the difference of the observed soft X-ray flux with respect
to an assumed model in the same band. A spectral break at $\sim
1-2$~keV may for this reason induce a high spurious $N_H$ measurement.

In the following sections we present the results of numerical
simulations for the evolution of the measured $N_H$ as a function of
the time elapsed after the onset of a source of ionizing photons in
its centre. We consider in particular the possibility of detecting a
variation of the column density during the $\sim 100$ seconds of
duration of a typical GRB. By imposing that the $N_H$ should remain
above the detection threshold for a reasonable time (1 s) but vary in
less that the GRB duration, we can strongly constrain the geometrical
properties of the material surrounding the GRB. We then compare our
results with the $N_H$ measurements performed in the early emission of
GRB~980329 (Frontera et al. 2000) and GRB~780506 (Connors \& Hueter
1998).

\section{Numerical Simulations}

\begin{figure}
\psfig{file=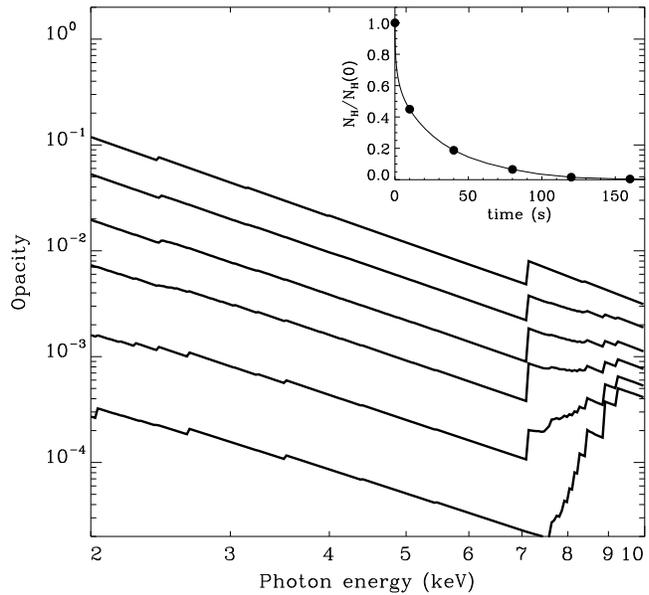,width=0.48\textwidth}
\caption{{Opacity in the range [2-10]~keV for a cloud with solar
metallicity, $R=3\times10^{18}$~cm and initial column density
$N_H(0)=3\times10^{21}$~cm$^{-2}$. In the main panel, from top to
bottom, we plot the absorption at times $t=0$, 10, 40, 80, 120 and 160
seconds. In the inset, the column density is shown as a function of
time. Filled dots underline the column densities for which a spectrum
is plotted in the main panel.}
\label{fig:nht}}
\end{figure}

In the case of a single atomic transition it is possible to derive
analytical equations to describe the time evolution of its opacity in
the optically thin regime (Lazzati et al. 2001c).  In the case of
continuum absorption the calculation is complicated by the need of
keeping track of all the ionization states of the various elements and
relative cross sections. For this reason the problem can be faced only
with the use of numerical simulations.

We have used the time-dependent photoionization code described in
Lazzati et al. (2001c), which is based on routines developed by
Raymond (1979) and specialized to the case of GRBs and their remnants
by Perna, Raymond \& Loeb (2000) and Perna \& Raymond (2000).  As
described by Perna et al. (2000), this code computes in a fully
time-dependent fashion the ionization state of the 12 most abundant
astrophysical elements, while updating, at each time step, the
temperature of the plasma\footnote{See also Schwarz 1973 for related
work.}.  The ionizing continuum is simulated in the range
[0.01-10]~keV with a spectrum relevant to GRB emission\footnote{Note
that the ionizing continuum has to be simulated in a spectral band
wider than the one in which the absorption is measured since soft
photons can ionize ions that contribute to the opacity at higher
frequencies.}. We adopted a photon index $\Gamma=1$ and constant
luminosity $L_X=1.3\times10^{49}$~erg~s$^{-1}$ in the [2--10] keV band
for $t\ge t_0$.  We considered three different radial distributions of
the density of absorbers: a uniform cloud with $R=3\times10^{18}$~cm,
a shell with $R=3\times10^{18}$~cm and $\Delta R/R=0.1$ and a wind
with $n(r)=n_0(R/R_{\min})^{-2}$ for $R\ge R_{\min}=10^{13}$~cm and
$n(r)=0$ for smaller radii.  For the uniform and shell geometries, we
performed simulations with initial column densities ranging from
$3\times10^{19}$~cm$^{-2}$ to $3\times10^{24}$~cm$^{-2}$,
logarithmically spaced by a factor $\sqrt{10}$. In the case of the
wind geometry, we considered only high values of the initial column
density because of its very rapid decrease with time, and we performed
simulations for the initial values $N_H(0) = 10^{22}$, $10^{23}$ and
$10^{24}$~cm$^{-2}$.

Even though the simulations for the uniform and shell geometries are
performed for a single value of the outer radius of the absorber
distribution (see above), these can be applied to rescaled matter
distributions.  In fact, given a distribution of absorbers $n(r)$ and
the ionizing luminosity $L_X$, the results of the simulations can be
extended to any distribution $n'(r) = a^{-1}\,n(r/a)$ (where $a$ is a
scale parameter) and any luminosity $L'_X=b\,L_X$.  The time of
photoionization of an ion is proportional to the flux, which scales as
$F\propto L/r^2$, where $r$ is the distance of the ion from the photon
source and $L$ the luminosity of the ionizing photons. For this reason
\begin{equation}
N_{H_{\{a,b\}}}(t) = N_H\left({{b\,t}\over{a^2}}\right)
\end{equation}
where $N_H(t)$ is the column density as a function of time from the
simulation and $N_{H_{\{a,b\}}}(t)$ is the column density as a
function of time for the ionizing luminosity $L'_X=b\,L_X$ and the 
distribution of absorbers $n'(r) = a^{-1}\,n(r/a)$.

\begin{figure}
\psfig{file=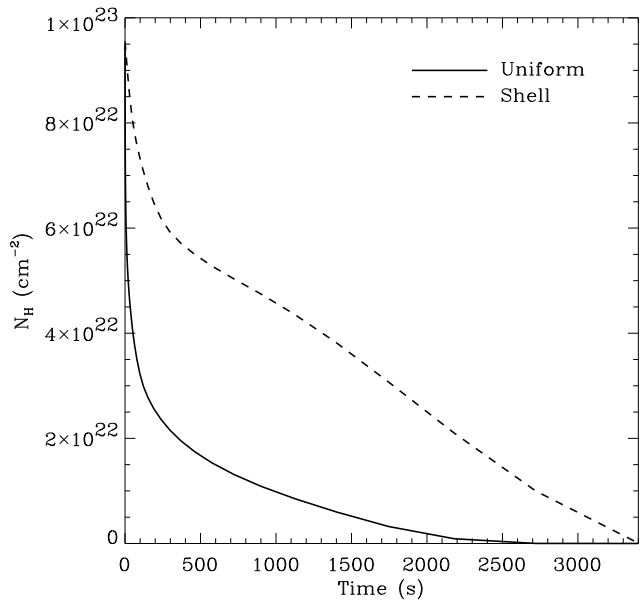,width=0.48\textwidth}
\caption{{Evolution of the column density with time for a uniform 
(solid line) and shell (dashed line) environments. In both cases the
initial column density is $N_H(0)=10^{23}$~cm$^{-2}$ and the outer
radius is $R=1$~pc.}
\label{fig:geo}}
\end{figure}

\subsection{Computation of the ``observed'' column density evolution}

The above described numerical code produces frequency and
time-dependent opacities $\sigma(\nu,t)$ by fully accounting for the
spectral variation with time of the opacity as the gas makes the
transition from neutral to fully ionized; as such, however, the output
of the code contains much more information than the observational
data.  This is particularly true for the GRB prompt X-ray emission,
usually detected with proportional counter detectors, such as the {\it
Beppo}SAX Wide Field Cameras (WFC). In particular, what is usually
measured from the observed spectra is the amount of soft X-ray
absorption, parameterized with the hydrogen column density in the
assumption that all the material is cold, i.e. neutral (Morrison \&
McCammon 1983). We hence applied an analogous procedure to our
synthetic dataset, averaging the opacity in the observed band in order
to reduce the information contained in the simulated spectrum to a
column density to be compared with the observations.  This is
equivalent to assume that the time and frequency-dependent opacity
can be decomposed as $\tau(\nu,t)=N_H(t)\,\sigma(\nu)$, where
$\sigma(\nu)$ is the average cross section at frequency $\nu$,
independent of time, weighted with the abundance of each element.  Our
procedure is also justified by the fact that the most widely used
software for X-ray data analysis (XSPEC; Arnaud 1996) neglects the
change of $\sigma(\nu)$ with ionization to compute the column density
(model WABS) when the single ion features cannot be
resolved\footnote{Note that there are cases, such as that of the warm
absorbers, where the effects of the spectral variations of the opacity
can actually be observed through the enhancement of the absorption edges of
intermediate $Z$ elements (see, e.g., Fabian et al. 1994 for a more
refined treatment of warm absorbers). In our case, however, the
situation is made more difficult by the fact that the absorber is not
in the same ionization state at a given time $t$. The inner regions
will be completely ionized at a time in which the outer regions are
still cold, while gas at intermediate distance is warm.}.  The
consequence of this assumption is that the column density at a given
time has a small dependence on the band in which it is measured.  In
the present paper we consider the energy range [2--10]~keV, since our
aim is to compare our results with the measurements performed with the
{\it Beppo}SAX WFC.

In the above approximation, we compute the column density as:
\begin{equation}
N_H(t) = N_H(0) \left\langle {{\tau(\nu,t)}\over{\tau(\nu,0)}} 
\right\rangle_{\rm[2-10]}
\end{equation}
where the symbol $\langle \rangle_{\rm[2-10]}$ represents the average
over the frequency range [2--10]~keV.  In Fig.~\ref{fig:nht} we show
the results of a simulation for a uniform cloud with
$N_H(0)=3\times10^{21}$~cm$^{-2}$ and $R=3\times10^{18}$~cm. The
opacities as a function of frequency are plotted for several times
after the burst onset, while the inset shows the evolution of the
column density in the first 100 s of the simulation.

In Fig.~\ref{fig:geo} we show the results of two simulations with
identical initial conditions, but for different geometries of the
absorbing material. The evolution of the absorbing column density in a
uniform cloud and a shell are shown with a solid and a dashed line,
respectively. The column density of a shell environment has a roughly
linear decay, slower than that of the uniform cloud. This is due to
the fact that in a shell geometry all the material is located at a
large distance from the source of ionizing photons, and hence the
ionization time is longer. With a good dataset, this difference can be
succesfully used to investigate the radial structure of the absorbing
medium.

\section{Burst environment}

\begin{figure}
\psfig{file=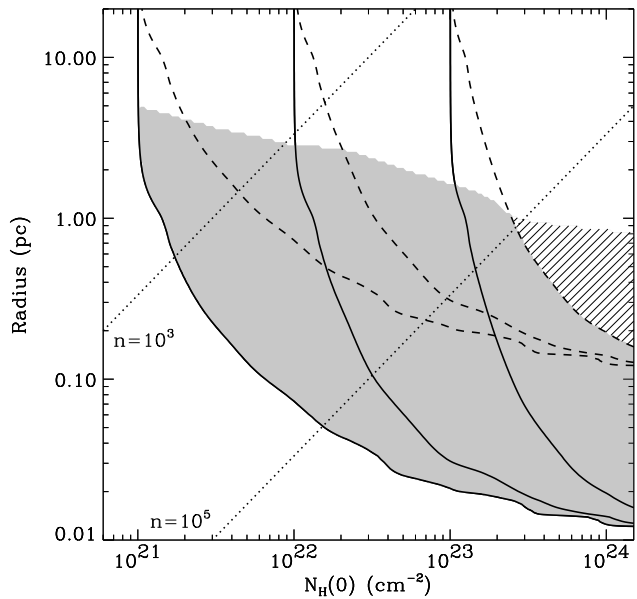,width=0.48\textwidth}
\caption{{Contour plot of $N_H(1)$ (solid contours) and $N_H(100)$ (dashed
contours) in the radius vs. column density parameter space. From lower
left to upper right, the solid contours show $N_H(1) = 10^{21}$,
$10^{22}$ and $10^{23}$~cm$^{-2}$ respectively. The same levels are
plotted as dashed lines for $N_H(100)$. The shaded region shows the
area in which both conditions expressed by Eqs. (\ref{eq:c1}) and
(\ref{eq:c2}) are satisfied. The line-shaded area shows the region in
which the [2--10] keV band is obscured for the whole burst duration
($N_H(100)\gsim 10^{23}$~cm$^{-2}$).}
\label{fig:uni}}
\end{figure}

In several bursts, detection of variable low-energy X-ray absorption
has been claimed (GRB~780506, Connors \& Hueter 1998; GRB~980329,
Frontera et al. 2000; GRB~990705, Amati et al. 2000; GRB~010222, in't
Zand et al. 2001). In this section we examine under which conditions
such an evolution of the measured column density can be detected
during the prompt emission of the bursts.

In order to detect an $N_H$ variation it is sufficient (but also
necessary) that the following two constraints be satisfied.  First,
the spectrum of the GRB must be integrated in order to perform a
measurement of column density. The minimum integration time (see
e.g. Frontera et al. 2000) is of the order of several seconds.
Variations of the column density can then be detected only if
\begin{equation}
N_H(1) > N_{H_{\rm Gal}} \sim 10^{21} \;\; {\rm cm}^{-2}\;,
\label{eq:c1}
\end{equation}
where $N_H(1)$ is the measured $N_H$ one second after the burst onset,
and $N_{H_{\rm Gal}}$ is the Galactic column density. With this
condition we ensure that the column density in the surroundings of the
GRB gives a detectable signal in the first integration bin.

On the other hand, in order to detect a variation, the column density
must sensibly decrease during the GRB emission time. Our second
condition is then that the observed column density decrease by a
factor of two in less than 100 seconds, the average burst
duration\footnote{The factor $2$ that we use in Eq.~\ref{eq:c2} has
been chosen since a sizable decrease of the absorbing column is needed
in order to have a significant measurement of its variation (see
below). We tested different values (namely 1.1 and 10) finding that
the allowed region changes slightly, but the conclusions are not
significantly affected.}:
\begin{equation}
N_H(100) \le N_H(1)/2.
\label{eq:c2}
\end{equation}

\begin{figure}
\psfig{file=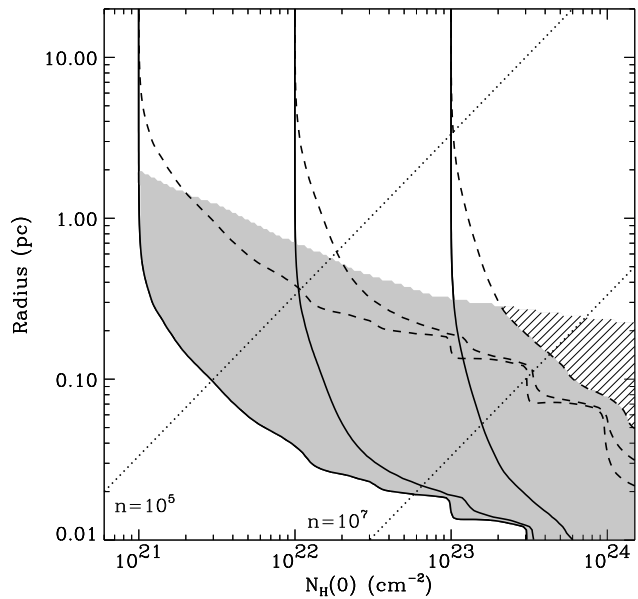,width=0.48\textwidth}
\caption{{Same as Fig.~\ref{fig:uni} but for a shell geometry. 
The shell width is assumed to be 10 per cent of its radius.}
\label{fig:she}}
\end{figure}

We must also consider that the burst itself has to be detected.
Since the soft $\gamma$-rays, in which GRBs are usually detected, are
not attenuated by photoabsorption, we consider Thomson scattering as
the most effective attenuation process. We then have
$N_H(0)\le1/\sigma_T\approx1.5\times10^{24}$~cm$^{-2}$. If
$N_H(0)\gsim 10^{23}$~cm$^{-2}$, the [2--10] keV flux is completely
absorbed and only a lower limit to the column density can be
measured. In the extreme case of a large and dense cloud (line-shaded
region in Fig~\ref{fig:uni}, \ref{fig:she} and \ref{fig:ger}) the
opacity in the [2--10]~keV range is larger than unity for the whole
duration of the bursts, and no flux can be detected in this band.

\begin{figure}
\psfig{file=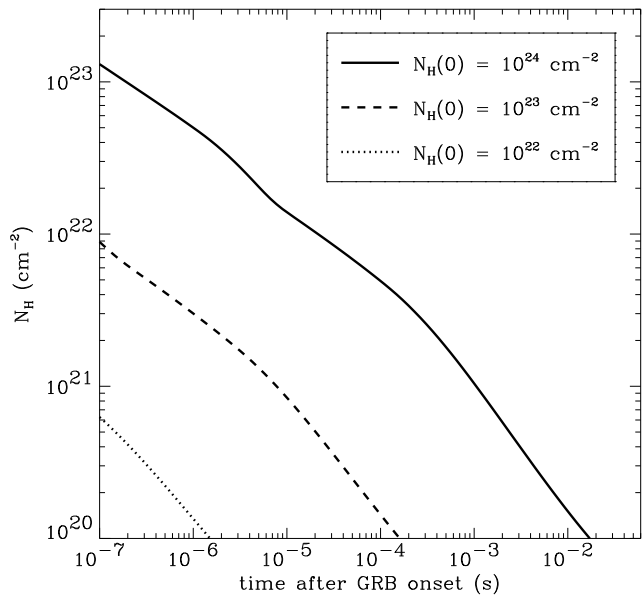,width=0.48\textwidth}
\caption{{Measured column density as a function of time in a wind
geometry with $R_{\min}=10^{13}$~cm. The three lines show different
initial column densities (see inset). In all cases, after a fraction
of a second the observable column density drops to values well below
the detection limit.}
\label{fig:wind}}
\end{figure}

\begin{figure}
\psfig{file=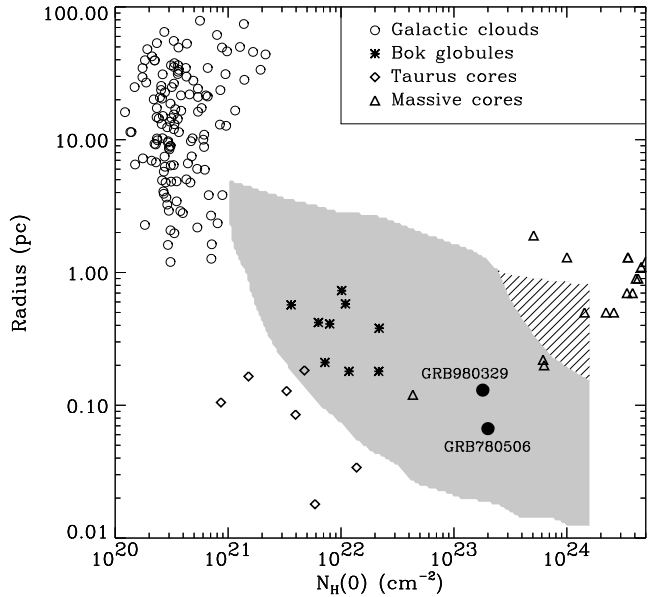,width=0.48\textwidth}
\caption{{Comparison between the radius and column densities inferred from 
Fig.~\ref{fig:uni} and the properties of Galactic molecular clouds and
some hierarchical structures embedded in them. The gray and line
shaded areas have the same meaning of Fig.~\ref{fig:uni}, i.e. the
region in which a column density variation should be observed during
the prompt GRB phase. Circles show radii and average column densities
of a sample of Galactic molecular clouds (Leisawitz et
al. 1989). Asterisks refer to Bok globules (Launhardt et al. 1998),
diamonds refer to dense cores in the Taurus molecular cloud (Zhou et
al. 1994) and triangles refer to massive cloud cores (Plume et
al. 1997). The two filled dots are the best fit values to the $N_H$
measurements of GRB~780506 and GRB~980329 (see Sect.~\ref{sec:fit1}
and~\ref{sec:fit2}).}
\label{fig:ger}}
\end{figure}

In Fig.~\ref{fig:uni}, \ref{fig:she} and \ref{fig:wind} we show that
conditions (\ref{eq:c1}) and (\ref{eq:c2}) can be satisfied only in a
limited range of cloud radii, geometries and initial column densities.
In Fig.~\ref{fig:uni} we show the column density at time $t=1$~s and
$t=100$~s as a function of the cloud radius and initial column density
with solid and dashed contours, respectively. The shaded area
underlines the region in which both conditions are satisfied.  This
shows that, in order to detect $N_H$ variations during the prompt
emission of the burst, the absorbing region surrounding the GRB must
be compact, with a radius $R<5$~pc and initial column density
$N_H(0)>10^{21}$~cm$^{-2}$. This translates to volume densities
$n\gsim 10^3$~cm$^{-3}$ (dotted lines). Figure~\ref{fig:she} shows the
same results but for a shell with $\Delta R /R =0.1$.

The possible contribution of a wind environment to the column density
evolution is considered in Fig.~\ref{fig:wind}. In the figure we plot
the measured $N_H$ as a function of time for three different initial
column densities: from a thin $N_H(0) = 10^{22}$~cm$^{-2}$ wind to a
maximum value of $N_H(0) = 10^{24}$~cm$^{-2}$. Figure~\ref{fig:wind}
shows how, no matter the initial value of the column density, it drops
to an undetectable level in less then a fraction of a second.
This is due to the fact that, in the case of a wind geometry,
the column density is dominated by the region very close to $R_{\rm min}$,
for which the ionization time is very short.

In order to compare the inferred cloud properties with the properties
of Galactic molecular clouds, in Fig.~\ref{fig:ger} we compare the
gray region in the parameter space with the properties of Galactic
molecular clouds and their denser regions. We assume that the burst
explodes in the centre of these regions. In Fig.~\ref{fig:ger}, the
circles show the properties of a sample of cloud radii and masses
presented in Leisawitz et al. (1989), from which we derived average
column densities.  There are no circles that fall in the allowed
region shaded in grey. Inside molecular clouds, however, there are
overdense regions, where star formation is supposed to take place.
The triangles show the position in the diagram of a sample of dense
molecular cloud cores, selected for their presence of H$_2$O masers
and in which massive star formation is taking place (Plume et
al. 1997). Their radii and column densities are consistent with the
shaded region. In some cases the column density is even too high.  A
sample of Bok globules (Launhardt et al. 1998), thought to be
directely related to star formation, is shown with asterisks. Their
properties are fully consistent with the shaded region. Finally, we
plot with diamonds the properties of dense cores in the Taurus
molecular clouds, where low-mass star formation is though to happen
(Zhou et al. 1994). These are too compact with respect to the shaded
regions. It is however worth mentioning that the properties of these
overdense regions at the time of the burst explosion may be somewhat
different, due to the radiation pressure exerted by the luminosity of
the massive star progenitor to the burst.

Should these $N_H$ variations be confirmed, they indicate
that the burst progenitors (or at least some of them) are located
inside overdense regions of molecular clouds. The properties of these
regions are consistent with those of the places in which massive star
formation takes place.

\subsection{GRB~980329}
\label{sec:fit1}

\begin{figure}
\psfig{file=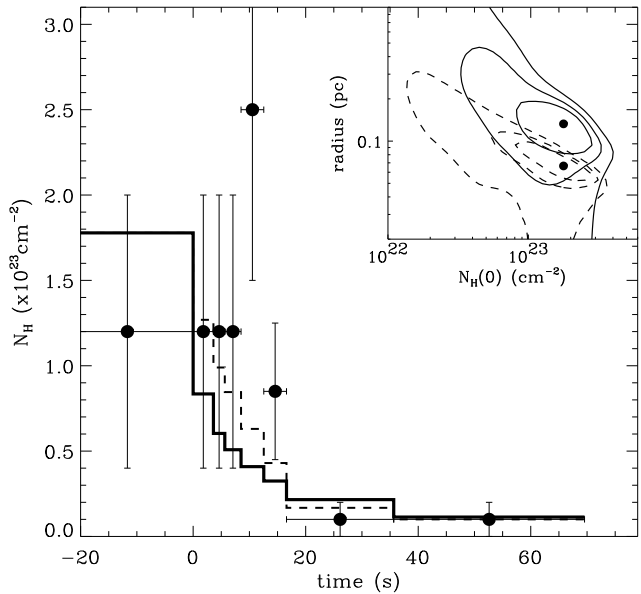,width=0.48\textwidth}
\caption{{Evolution of the column density measured in the prompt X-ray
spectrum of GRB~980329 (Frontera et al. 2000). A uniform and shell
geometry for the absorbing column have been tested. The best fit
models, integrated over the same time intervals of teh data, are shown
with a solid and a dashed line, for the uniform and shell geometries,
respectively. The inset shows the $1\sigma$, 90 per cent and 99 per
cent confidence contours for the two fitted parameters. Again, solid
contours refer to uniform density and dashed contuors to a shell
geometry.}
\label{fig:980}}
\end{figure}

To date, three GRBs detected by the WFC on board {\it Beppo}SAX showed
evidence of variable soft X-ray absorption. Published data are
available for GRB~980329 (Frontera et al. 2000). These data are shown,
along with their uncertainties, in Fig.~\ref{fig:980}. The error bars
on the last two measurements, non quoted in the original paper, have
been assumed equal to the measurements themselves. The timescale
has been reported to the beginning of the flaring activity of the
burst, which starts $\sim 23$~s after the trigger (see Fig. 1 of
Frontera et al. 2000)

We fitted our uniform-cloud and shell models for the evolution of the
column density to the data. For the uniform model we obtained the best
values $N_H(0)=1.8\times10^{23}$~cm$^{-2}$ and $r=0.13$~pc
($n\sim4.5\times10^5$~cm$^{-3}$), while in the shell case we obtain
$N_H(0)=1.8\times10^{23}$~cm$^{-2}$ and $r=0.066$~pc. In
Fig.~\ref{fig:980}, the best fit models are shown with a solid line
(uniform) and dashed line (shell). The shell model has a better
$\chi^2$ value, but due to the large uncertainties in the measurement,
the two geometries cannot be significantly disentangled. The upper
right inset shows the $1\sigma$, 90 per cent and 99 per cent
confidence contours for the two parameters in both cases.  Since the
statistical significance of the variation of the column density is not
compelling, the 99 per cent confidence contour is not closed.  The
different evolution of the two models shows, however, that with good
quality data it is in principle possible to understand what is the
geometry of the medium surrounding GRBs.

A problem in the fit of GRB~980329 data is the lack of a redshift
measurement of the burst. In fact all our simulations are performed in
the rest frame. The model should be however modified in two ways if
the source of absorption lies at redshift $z>0$.  First, the timescale
is stretched by a factor $(1+z)$. Most importantly, if a column density
$N_{H_z}$ surrounds a source at redshift $z$, the soft X-ray
absorption measurement will yield a measured value
$N_H=(1+z)^{-2.5}\,N_{H_z}$, where the exponent $-2.5$ holds in the
redshift range $0<z<4$.  This implies that for a source at redshift
$z=1$ the measured column density is underestimated by a factor $\sim
5$. The modelling of the data with a known redshift may hence give a
best fit which is an even more compact and dense cloud.

\subsection{GRB~780506}
\label{sec:fit2}

\begin{figure}
\psfig{file=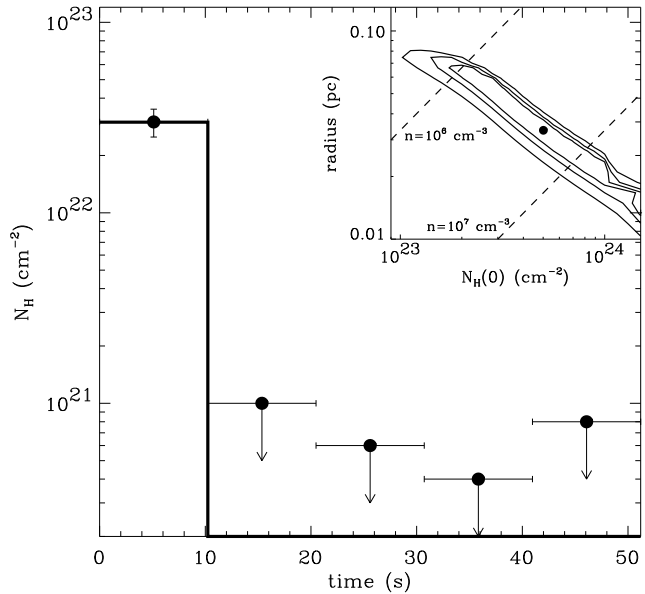,width=0.48\textwidth}
\caption{{Same as Fig.~\ref{fig:980} but for GRB~780506 (Connors \& 
Hueter 1998). The solid line shows the best fit model with $N_H(0)=
2\times10^{23}$~cm$^{-2}$ and $R=2\times10^{17}$~cm integrated over
the observed time intervals. The dashed line shows the best fit model
without the time integration. The inset shows the $1\sigma$, 90 per
cent and 99 per cent contour levels for the two parameters and, with
dashed lines, isodensity contours for $n=10^6$ and $10^7$~cm$^{-3}$.}
\label{fig:780}}
\end{figure}

Besides the three bursts detected by the {\it Beppo}SAX WFC, a
measurement of column density was possible in a burst serendipitously
detected in 1987 by the HEAO-1 satellite with the A-4 and A-2
instruments (GRB~780506, Connors \& Hueter 1998). The burst spectrum
was fitted with a three component model: a power-law, a thermal
component and an absorption component, with the column density free to
vary. The lightcurve was integrated in time bins of 10.24 s each and
the six spectra were fitted indipendently. The measured $N_H$ values
are shown in Fig.~\ref{fig:780}.  In the first time bin a very
significant ($6\sigma$) soft X-ray absorption was detected, while the
fits in the remaining 5 time intervals yielded only upper
limits. Fitting our uniform cloud model to the data, again without the
possibility to correct for the redshift of the burster, we find a best
fit cloud with $R=2\times10^{17}$~cm and
$N_H(0)=2\times10^{23}$~cm$^{-2}$.  Since only one positive
measurement was possible, as shown by the inset in Fig.~\ref{fig:780},
the radius and initial column density of the cloud cannot be precisely
determined simultaneously. We can safely conclude, however, that the
density of the cloud was $n\gsim10^6$~cm$^{-3}$.  Since for this burst
only one positive measurement was obtained, it is not possible to
distinguish between a uniform and a shell geometry.

\section{Discussion and Conclusions}

We have performed numerical simulations to compute the evolution of
the soft X-ray absorption of a cloud under the ionization flux of a
Gamma-Ray Burst exploded at its centre. We have considered a uniform
cloud, a shell and a wind geometry, and we have shown that variations
of the column density during the duration of a GRB can be detected
only under well-defined geometrical conditions of the absorbing
material.

The column density of a wind, in which the bulk of the material
responsible for the absorption is close to the GRB, cannot survive the
first milliseconds of the illumination and is undetectable with the
present instrumentation. On the other hand, if the material is spread
uniformly over a region with size $0.05 \lsim R\lsim 5$~pc, the soft
X-ray absorption can be observed to evolve in time.  Present data do
not allow to firmly detect the evolution of the column density,
nor to tell if the material is uniformly spread around the source or
rather concentrated in a shell at distance $R$ from the burster.  By
comparing the allowed properties of molecular clouds with the results
presented here,  we conclude that the weak evidence of $N_H$
variation during the prompt X-ray emission of some GRBs can be
accounted for if they are associated to compact overdense regions
similar to the cocoon of star formation within molecular clouds or to
Bok globules.

A number of approximations have been applied to this analysis and are
worth a discussion. First, we consider an ionizing flux with fixed
spectrum and luminosity. Both these assumptions are coarse
approximations of a GRB lightcurve and spectral evolution. Even though
GRB are frantically variable, the results of our simulations are exact
if the time $t$ is substituted with the quantity $t' = \int_0^t L(t)\,dt/
\int_0^{t_{\rm GRB}} L(t)\,dt$, where $t_{\rm GRB}$ is the total burst 
duration. Since in this work the detailed time variability of the
$N_H$ lightcurve is not addressed, the approximation of constant
luminosity is reasonable.

A more important issue is the variation of the spectral slope with
time.  More complete simulations of time-dependent column densities
will be presented in a forthcoming paper (Perna et al. in
preparation), in which the dependence of the observed column density
on the ionizing spectrum will be discussed in detail. However,
$\Gamma=1$ is a good value for the average time integrated low-energy
slope of GRBs (Preece et al. 2000).

We have assumed that the medium is at rest during the photoionization
process.  While Lazzati et al. (2001c) showed that the fireball
interaction is not important, radiation pressure may accelerate the
medium. The ratio of the timescale of photoionization $t_{\rm pi}$
over the timescale of acceleration $t_{\rm acc}$ is given, for a
$\Gamma=1$ spectrum, by:
\begin{equation}
{{t_{\rm pi}}\over{t_{\rm acc}}} = {{h\nu_{\rm b}\,\sigma_T}\over
{m_p\,c^2\,\sigma_{\rm pi}}} \sim 2\times10^{-6} \,
\left({{h\nu_b}\over{511\,{\rm keV}}}\right) \,
\left({{h\nu_{\rm th}}\over{1\,{\rm keV}}}\right)^{2.3}
\end{equation}
where $\nu_b$ is the break energy of the GRB spectrum and $\sigma_{\rm
pi}\sim2\times10^{-22} (h\nu_{\rm th}/1\,{\rm keV})^{-2.3}$ is the
average photoionization cross section at the threshold energy
$h\nu_{\rm th}$ weighted with the element solar abundance (Anders \&
Ebihara 1982). Even for iron, the relevant element with the highest
threshold energy, the acceleration timescale is several orders of
magnitude longer than the photoionization one.

Finally, as discussed in \S 4.1, the redshift of the burst introduces
time dilation and decreases the observed column density at a given
time. All these factors have not been taken into account in
Fig.~\ref{fig:uni} and Fig.~\ref{fig:she}, so that the gray shaded
regions should be considered indicative, with uncertainties up to a
factor of order unity. Nevertheless we can firmly conclude that if a
column density variation is observed in the first $\sim 100$ seconds
of the prompt emission of a GRB, then it must be located within a
cloud with properties roughly consistent with those of
Fig.~\ref{fig:uni} or inside a shell (Fig.~\ref{fig:she}).  We
have not considered here the possibility of a beamed fireball. In the
case of a conical jet, as discussed to date for GRB fireballs (Sari et
al. 1999, Ghisellini \& Lazzati 1999, Panaitescu \& Kumar 2001),
beaming does not affect our calculations, since all processes take
place along the line of sight. In the case of a cylindrical jet, the
radial dilution of photons would be different, and our computations
are not appplicable.

Variations in the soft X-ray absorption have indeed been claimed in
several bursts. Only one of them, however, shows this evidence quite
clearly. GRB~780506, serendipitously observed by the HEAO-1 satellite
(Connors \& Hueter 1998) has a highly significant soft X-ray
absorption in the first time integration bin, which vanishes in the
rest of the burst lightcurve (Fig.~\ref{fig:780}).  GRB~980329
(Frontera et al. 2000), shows a very well defined trend of the $N_H$
to decrease with time, which can be fit in our model (see
Fig.~\ref{fig:980}), but with large statistical uncertainties.
GRB~990705 (Amati et al. 2000) shows evidence of time dependent iron
absorption features. An evolution of the column density is also
claimed, but the data are not published. Finally, GRB~010222 (in 't
Zand et al. 2001) shows clear evidence of $N_H$ variations, but the
column density appears to rise and fall, a behavior that cannot be
explained in our model and is likely due to the presence of a spectral
break that moves inside and outside the observing window.

As a consequence of the high density ($n\ge10^3$) described above, the
afterglow of the bursts should evolve more rapidly than in a very low
density environment. In particular, the time at which the relativistic
to non-relativistic transition takes place is $t_{\rm
nr}\sim10\,(E_{53}/n_6)^{1/3}$~d (see, e.g. Wijers, Rees \& Meszaros
1997), where $E_{53}$ is the isotropic fireball energy in units of
$10^{53}$~erg and $n_6$ the density of the surrounding medium in units
of $10^6$~cm$^{-3}$. Unfortunately, there are not accurate afterglow
lightcurves for any of the bursts described in this paper. Indeed, a
very high density of the surrounding medium is required in GRB~990705
(Lazzati et al. 2001a). In the case of GRB~010222, which displays
$N_H$ variations but cannot be straightforwardly included in our
model, a very high density of the surrounding medium is claimed (in 't
Zand et al. 2001) on the base of spectral and lightcurve
considerations.

The HETE2 satellite has on-board two experiment, the Soft X-ray
Camera (SXC) and the Wide Field X-ray Monitor (WXM) potentially
sensitive to $N_H$ variations. The SXC, has an effective area of
$\sim10$~cm$^2$, is sensitive in the [0.5--10]~keV range and has a
field of view of 0.91~sr. The WXM has a larger effective area (175
cm$^2$) and field of view ($1.6$~sr) but is sensitive in the harder
[2--25]~keV band, where continuum absorption is less severe. The
combination of these two instruments, however, will hopefully clarify
the properties and time variability of the soft X-ray absorption in
the early prompt X-ray spectra of GRBs.

\section*{Acknowledgments}
We wish to thank the anonymous referee for his/her careful reading of
the manuscript and useful comments. We thank Gabriele Ghisellini,
Enrico Ramirez-Ruiz, Martin Rees, Luigi Stella and Mario Vietri for
very useful discussions and suggestions. We thank John Raymond for his
help in the process of updating the numerical code used in this work.

\end{document}